\documentclass{jps-cp}
\usepackage{txfonts} 

\title{Latest vertex and tracking detector developments for the future Electron-Ion Collider}

\author{Xuan \textsc{Li}$^{1}$}

\inst{$^{1}$ Physics Division, Los Alamos National Laboratory, Los Alamos, New Mexico, 87545, USA}

\email{xuanli@lanl.gov}

\recdate{November 30, 2022}

\abst{The high-luminosity high-energy Electron-Ion Collider (EIC) to be built at Brookhaven National Laboratory (BNL) will provide a clean environment to study several fundamental questions in the high energy and nuclear physics fields. A high granularity and low material budget vertex and tracking detector is required to provide precise measurements of primary and displaced vertex and track reconstruction with good momentum and spatial resolutions. The reference design of the EIC project detector at the first Interaction Point (IP), which was selected from three submitted EIC detector proposals, enables a series of high precision heavy flavor measurements. Based on the EIC project detector reference design, the newly formed ePIC collaboration is developing the technical design for the EIC project detector towards its construction and operation. The reference design of the EIC vertex and tracking detector consists of the Monolithic Active Pixel Sensor (MAPS) based silicon vertex and tracking subsystem, the Micro-Pattern Gas Detector (MPGD) based gas tracking subsystem and the AC-Coupled Low Gain Avalanche Diode (AC-LGAD) based silicon outer tracker, and it has the track reconstruction capability in the pseudorapidity region of -3.5 to 3.5 with full azimuthal coverage. Further detector geometry optimization with a new magnet and technology down selection are underway by the ePIC collaboration. The latest ePIC vertex and tracking detector geometry and its performance evaluated in simulation will be presented. Details of the EIC vertex and tracking detector R$\&$D, which include the proposed detector technologies, prototype sensor characterization and detector mechanical design will be discussed as well.}

\kword{Electron-Ion Collider, silicon, MAPS, AC-LGAD, vertex, track, radiation}

\begin{document}
\maketitle

\section{Introduction}

The Electron-Ion Collider (EIC) will utilize high-luminosity high-energy electron+proton ($e+p$) and electron+nucleus ($e+A$) collisions to precisely study the nucleon/nucleus 3D structure, help address the proton spin puzzle, search for the gluon saturation and explore how quarks and gluons form visible matter inside vacuum and a nuclear medium \cite{eic_YR}. The EIC project has received CD1 stage approval from the US Department of Energy (DOE) in 2021, which officially announces the starting of the defined EIC project. At the EIC, the beam energies of proton and nucleus (mass number A = 2-238) are 41~GeV, 100~GeV to 275~GeV and the electron beam energies vary from 2.5~GeV to 18~GeV. The instantaneous luminosity at the EIC is around $10^{33-34} cm^{-2} s^{-1}$, which means the annual EIC operation can achieve around 10-100 $fb^{-1}$ integrated luminosities. The EIC bunch crossing rate is around 10~ns. Two detectors have been planned to be built at the EIC, the first one is the EIC project detector located at the 6 clock position of the accelerator and the second one is currently named as detector II, which will be located at the 8 clock position of the accelerator. Recent developments of the EIC project detector, which include the reference detector design, vertex and tracking subsystem performance, detector geometry optimization and related R$\&$D progress will be discussed.

\section{EIC project detector reference design and performance}

Since the EIC project achieved CD0 phase approval in 2020, three proto-collaborations: ECCE \cite{ecce}, ATHENA \cite{athena} and CORE \cite{core} have performed dedicated detector and physics studies for the EIC project detector design. The recommendation of the EIC detector review committee suggests that the ECCE detector design \cite{ecce} is selected as the EIC project detector reference design considering multiple factors such as detector performance, cost, risk and schedule.

\subsection{EIC project detector reference design}

\begin{figure}[tbh]
\includegraphics[width=0.44\textwidth]{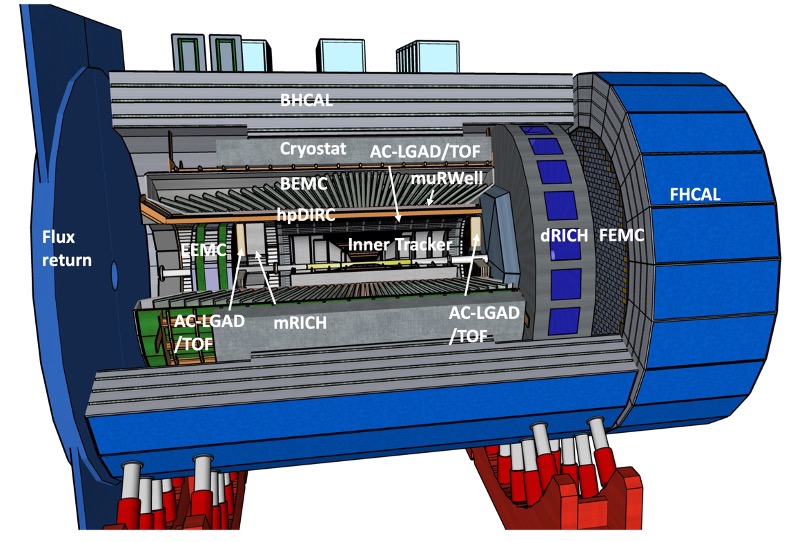}
\includegraphics[width=0.55\textwidth]{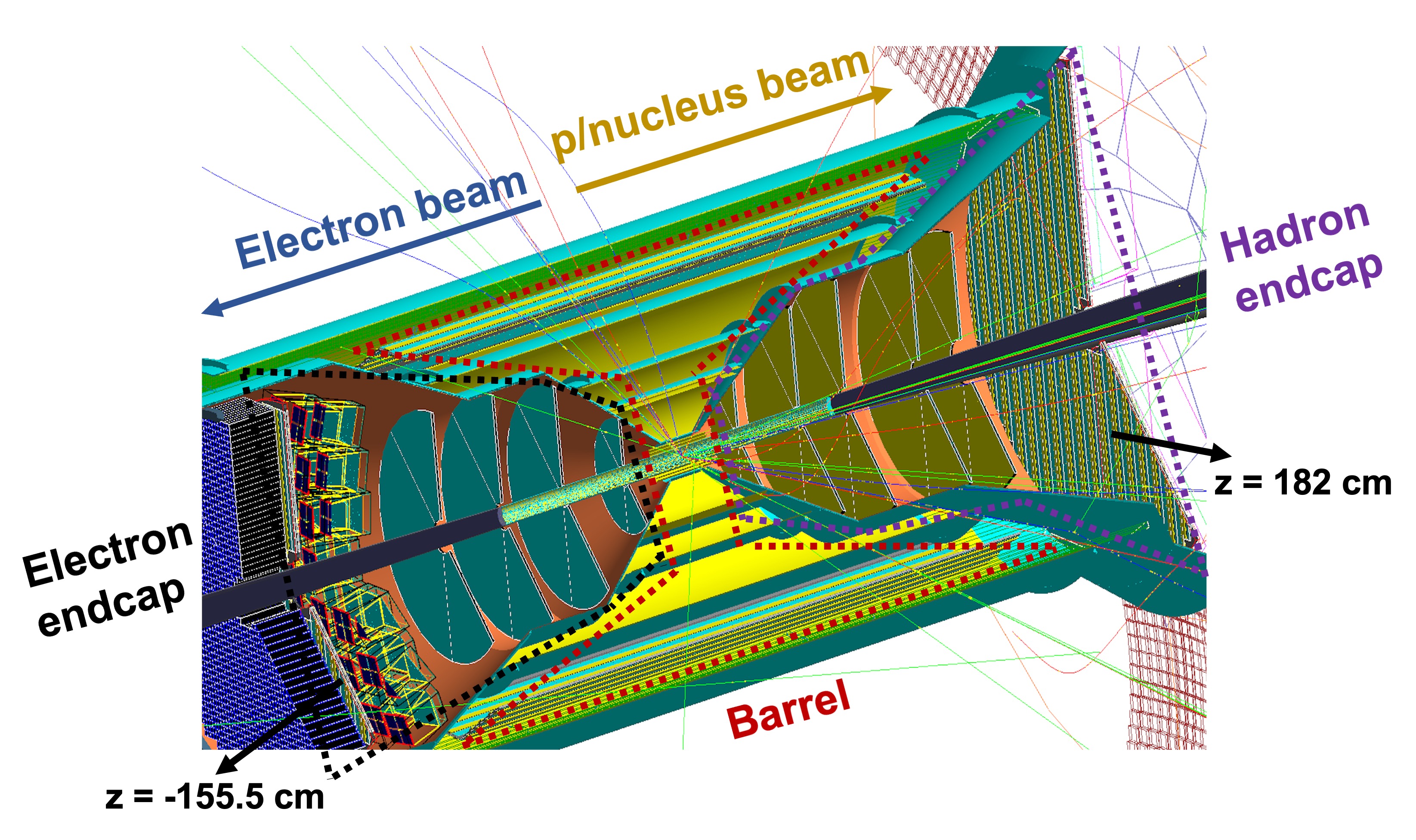}
\caption{Geometry of the EIC project detector reference design implemented in GEANT4 \cite{geant4} simulation (left) and the geometry of the vertex and tracking detector of the EIC project detector reference design (right). The left part of the detector locates in the electron beam going direction and the right part is in the proton/nucleus going direction. Detailed geometry parameters are listed in Table~\ref{ref_trk_barrel}, Table~\ref{ref_trk_had}, and Table~\ref{ref_trk_ele}.}
\label{ref_design}
\end{figure}

The selected EIC project detector reference design consists of a Monolithic Active Pixel Sensor (MAPS) \cite{maps} based silicon vertex and tracking detector, a Micro Pattern Gas Detector (MPGD) \cite{mpgd} based tracking detector, an AC coupled Low Gain Avalanche Diode (AC-LGAD) based Time of Flight (ToF) detector, a dual Ring-imaging Cherenkov detector (dRICH), a mirror Ring-imaging Cherenkov detector (mRICH), a Detector of Internally Reflected Cherenkov light (DIRC) PID detector, ElectroMagnetic Calorimeters (EMCal) and Hadronic Calorimeters (HCAL). This proposed detector reference design utilizes the existing Babar magnet with a maximum magnetic field at 1.4~T. It can provide precise primary and displaced vertex determination, tracking reconstruction, particle identification and energy measurements in the pseudorapidity region of $-3.5 < \eta < 3.5$. The layout of the EIC project detector reference design is shown in the left panel of Fig.~\ref{ref_design}.

To realize the proposed physics measurements, the EIC vertex and tracking detector is required to provide fine spatial resolution for primary and displaced vertex determination; low material budgets to mitigate multiple particle scattering to improve the track momentum resolution; sufficient number of hits for track pattern recognition finding; and fast timing to suppress backgrounds from the beam and neighboring collisions. A hybrid vertex and tracking detector option has been adapted by the EIC detector reference design, which consists of the next generation MAPS \cite{its3}, $\mu$RWELL \cite{muwell} and AC-LGAD based layers/disks. The high granularity thin MAPS vertex and tracking detector will provide better than 10 $\mu$m spatial resolution per hit, which is critical for precise vertex and track reconstruction. The $\mu$RWELL and AC-LGAD detectors will provide the large lever arm constraints and additional hits to improve the track finding quality. Moreover, fast timing capabilities provided by the AC-LGAD and $\mu$RWELL detectors will help pin down the combinatorial backgrounds for track reconstruction at EIC. The geometry of the EIC vertex and tracking detector is illustrated in the right panel of Fig.~\ref{ref_design}. According to the detector kinematic coverage, the EIC project detector vertex and tracking detector can be further divided into three subsystems: barrel, hadron endcap and electron endcap vertex and tracking detectors. Detailed geometry parameters of the EIC project detector vertex and tracking subsystem are shown in Table~\ref{ref_trk_barrel}, Table~\ref{ref_trk_had} and Table~\ref{ref_trk_ele}.

\begin{table}[tbh]
\caption{EIC project detector barrel vertex and tracking detector geometry}
\label{ref_trk_barrel}
\centering
\begin{tabular}{|clclc|c|c|c|}
\hline
Technology & Index &  $r$ (cm) & $z_{min}$ (cm) & $z_{max}$ (cm) & $X/X_{0}$  \\  \hline
MAPS    & 1  & 3.3 & -13.5 & 13.5 & 0.05\%   \\ 
MAPS    &2 & 4.35 & -13.5 & 13.5 & 0.05\%  \\ 
MAPS    &3 & 5.4 & -13.5 & 13.5 & 0.05\%  \\ 
MAPS    &4 & 21.0 & -27 & 27 & 0.24\%  \\ 
MAPS    &5 & 22.68 & -30 & 30  & 0.24\% \\ 
$\mu$RWELL  & 6 & 33.14 & -40 & 40 & 0.3\% \\ 
$\mu$RWELL  & 7 &  51 & -106 & 106 & 0.3\% \\
AC-LGAD     & 8 & 64 & -140 & 140 & 1\% \\ 
$\mu$RWELL  & 9 & 77.0 & -197 & 145 & 0.3\% \\
\hline
\end{tabular}
\end{table}

\begin{table}[tbh]
\caption{EIC project detector hadron endcap vertex and tracking detector geometry}
\label{ref_trk_had}
\centering
\begin{tabular}{|clclc|c|c|c|}
\hline
Technology & Index & $z$ (cm) & $r_{\rm in}$ (cm) & $r_{\rm out}$ (cm) & $X/X_{0}$  \\  \hline
MAPS    & 1  & 25 & 3.5 & 18.5 & 0.24\% \\ 
MAPS    & 2  & 49 & 3.5 & 36.5 & 0.24\% \\
MAPS    & 3  & 73 & 4.5 & 40.5 & 0.24\% \\ 
MAPS    & 4 & 106 & 5.5 & 41.5 & 0.24\% \\ 
MAPS    & 5 & 125 & 7.5 & 43.5  & 0.24\% \\
AC-LGAD  & 6  & 182 & 7 & 87  & $\sim$ 7\% \\ 
\hline
\end{tabular}
\end{table}

\begin{table}[tbh]
\caption{EIC project detector electron endcap vertex and tracking detector geometry}
\label{ref_trk_ele}
\centering
\begin{tabular}{|clclc|c|c|c|}
\hline
Technology & Index & $z$ (cm) & $r_{\rm in}$ (cm) & $r_{\rm out}$ (cm) & $X/X_{0}$  \\  \hline
MAPS    & 1  & -25 & 3.5 & 18.5 & 0.24\% \\ 
MAPS    & 2 &  -52 & 3.5 & 36.5 & 0.24\% \\ 
MAPS    & 3 &   -79 & 4.5 & 40.5 & 0.24\% \\ 
MAPS    & 4 &  -106 & 5.5 & 41.5 & 0.24\% \\ 
AC-LGAD  & 5  & -155.5 & 8 & 64 & $\sim$ 6\% \\
\hline
\end{tabular}
\end{table}

\subsection{EIC project detector reference design performance}

\begin{figure}[hbt]
\centering
\includegraphics[width=0.98\textwidth]{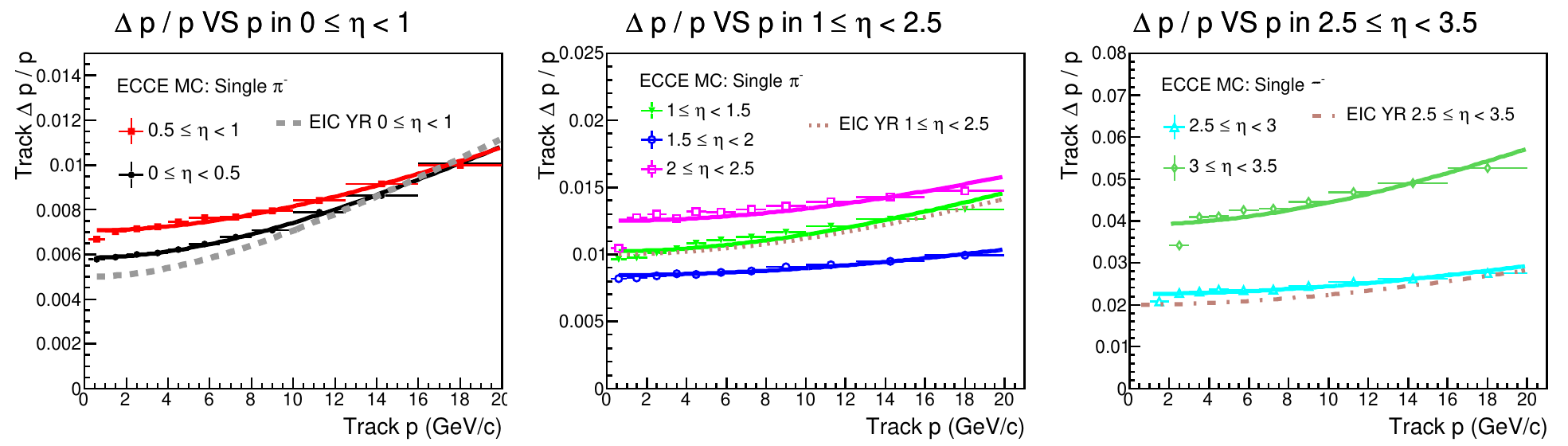}
\includegraphics[width=0.98\textwidth]{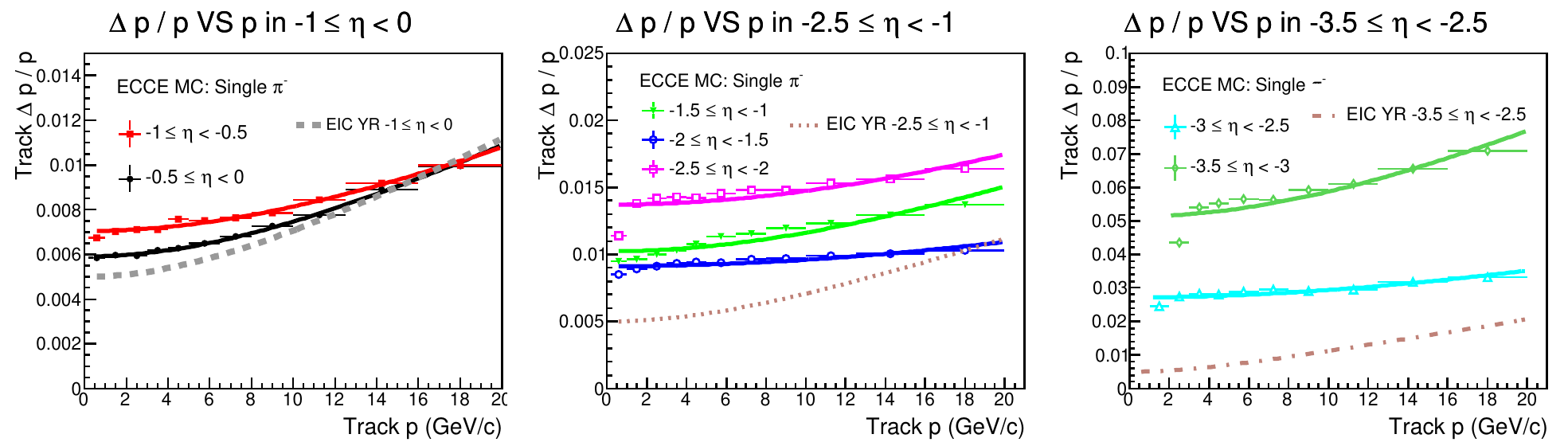}
\caption{Track momentum dependent momentum resolution of the EIC project detector reference design in the pseudorapidity regions of $0 \le |\eta| < 1$, $1 \le |\eta| < 2.5$ and $2.5 \le |\eta| < 3.5$. The tracking performance is evaluated with the 1.4T Babar magnet. The EIC yellow report tracking requirements in the respective pseudorapidity regions are highlighted in brown dashed lines.}
\label{ref_trk_mom}
\end{figure}

\begin{figure}[hbt]
\centering
\includegraphics[width=0.98\textwidth]{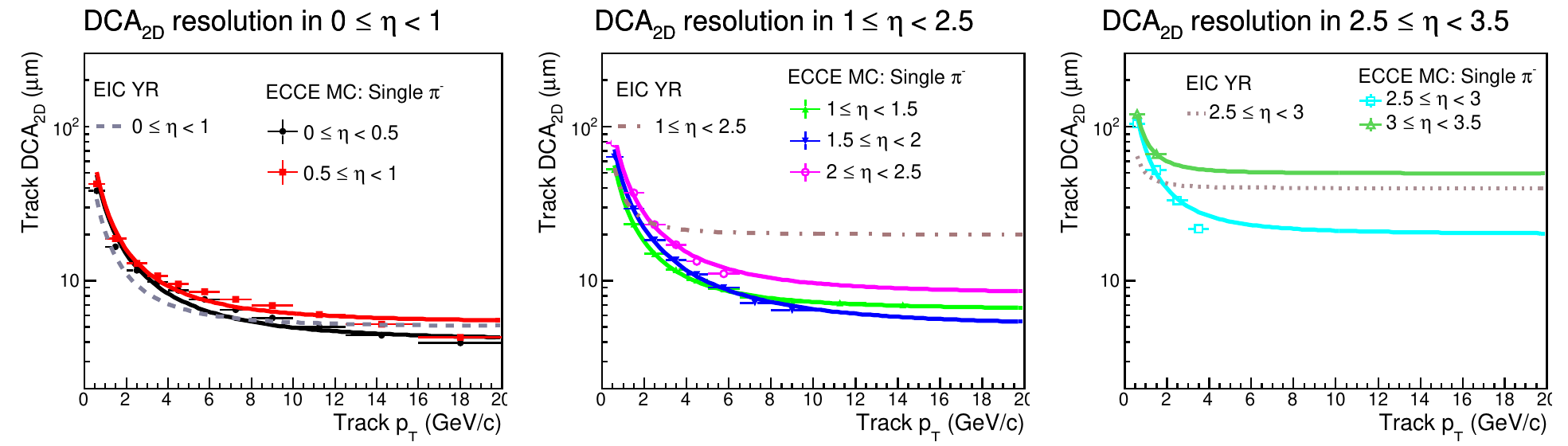}
\includegraphics[width=0.98\textwidth]{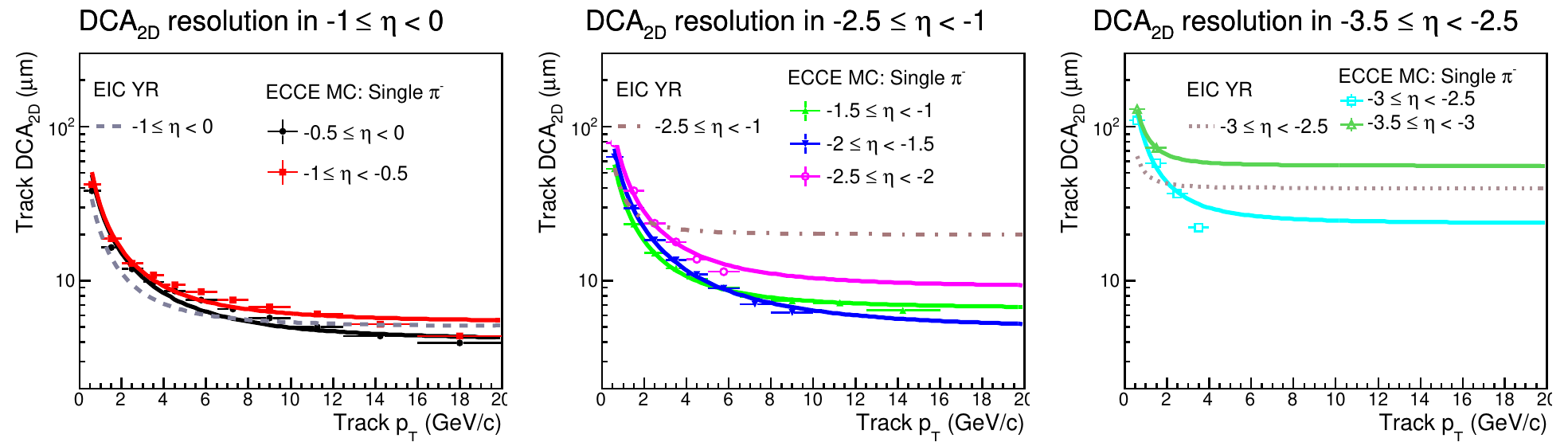}
\caption{Track transverse momentum dependent transverse Distance of Closest Approach ($DCA_{2D}$) resolution of the EIC project detector reference design in the pseudorapidity regions of $0 \le |\eta| < 1$, $1 \le |\eta| < 2.5$ and $2.5 \le |\eta| < 3.5$. The tracking performance is evaluated with the 1.4T Babar magnet. The EIC yellow report tracking requirements in the respective pseudorapidity regions are highlighted in brown dashed lines.}
\label{ref_trk_dca}
\end{figure}

\begin{figure}[hbt!]
\centering
\includegraphics[width=0.9\textwidth]{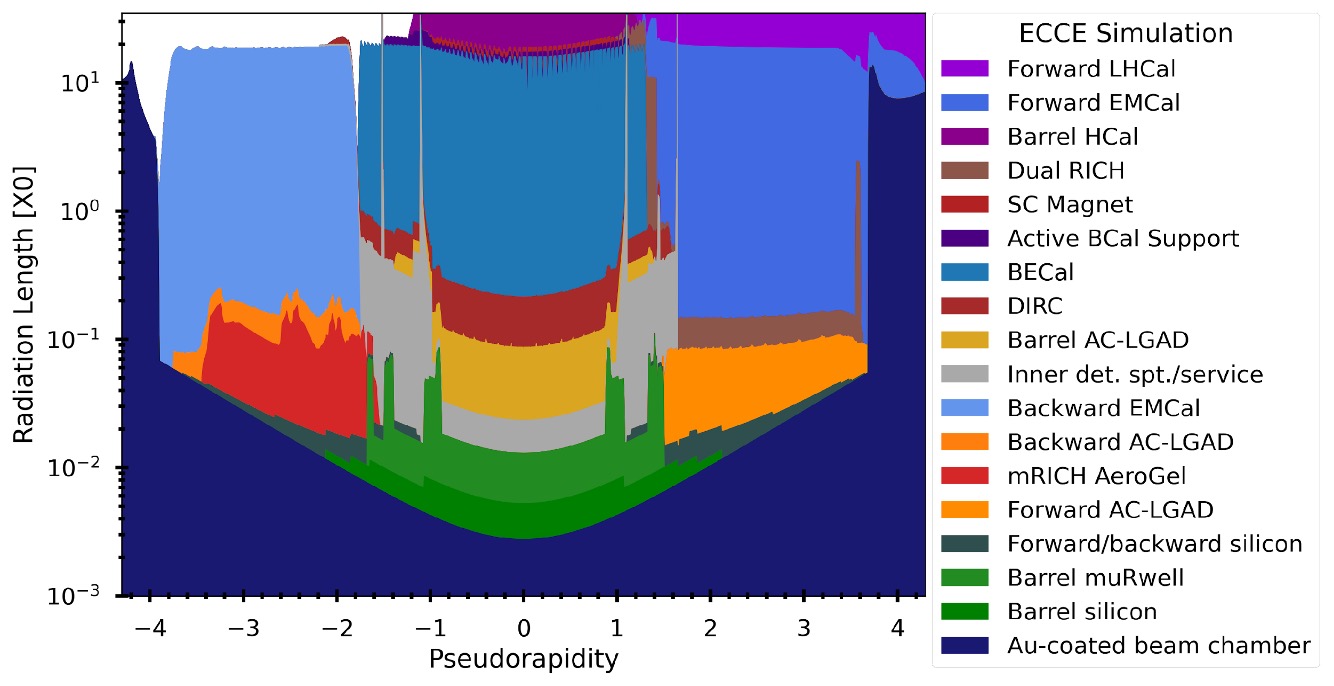}
\caption{Pseudorapidity dependent material budget scan of the EIC project detector reference design in simulation \cite{ecce}. In addition to active detector volume, material budgets of a service cone which consists of the support structure, cooling components and service parts for the inner vertex and tracking detector have been included as well.}
\label{ref_mat_scan}
\end{figure}

Geometry of the EIC project detector reference design has been implemented in GEANT4 \cite{geant4} simulation. In addition to detector active volume, the support structure, cooling components, power and data cables have been implemented in simulation as well. Tracking performance in different pseudorapidity regions has been evaluated with the EIC project detector reference design and the Babar magnet field map in simulation. Fig.~\ref{ref_trk_mom} presents the track momentum dependent momentum resolution and Fig.\ref{ref_trk_dca} shows the transverse momentum dependent transverse Distance of Closest Approach ($DCA_{2D}$) resolution in the pseudorapidity regions of $0 \le |\eta| < 1$, $1 \le |\eta| < 2.5$ and $2.5 \le |\eta| < 3.5$ using the EIC project detector reference design with single charged particle simulation. Precise electron tagging is desired for Deeply Inelastic Scattering (DIS) measurements at EIC, therefore the EIC yellow report has a more stringent detector requirement on tracking in the electron beam going direction (the negative pseudorapidity region) than the hadron nucleon/nucleus beam going direction (the positive pseudorapidity region). Compared to the EIC yellow report detector requirements \cite{eic_YR}, the tracking performance of the EIC project detector reference design is not far away from the desired performance in the pseudorapidity region of $0 \le |\eta| < 2.5$ especially in the high momentum or high transverse momentum region. However, tracking performance in the pseudorapidity region of $2.5 \le |\eta| < 3.5$ needs improvement to align with the EIC yellow report requirements. Similar features happen in the negative pseudorapidity regions as well \cite{ecce}.

Material budgets of the EIC project detector reference design have been scanned in simulation. Fig.~\ref{ref_mat_scan} shows pseudorapidity dependent material budget for different detector subsystems of the EIC project detector reference design. Although the total material budgets of the MAPS based silicon vertex and tracking detector and the MPGD based tracking detector are below 5$\%$ radiation length per the EIC yellow report requirements, the material budgets of the associated service cone are around 10-15$\%$ radiation length in the psedurapidity region of $1<|\eta|<1.6$. These non-negligible material budget of the service cone will impact on the tracking performance in the corresponding kinematic regions and further studies, which include detector geometry optimization and service cone routing optimization, are needed to improve the associated tracking performance.

\begin{figure}[tbh]
\centering
\includegraphics[width=0.74\textwidth]{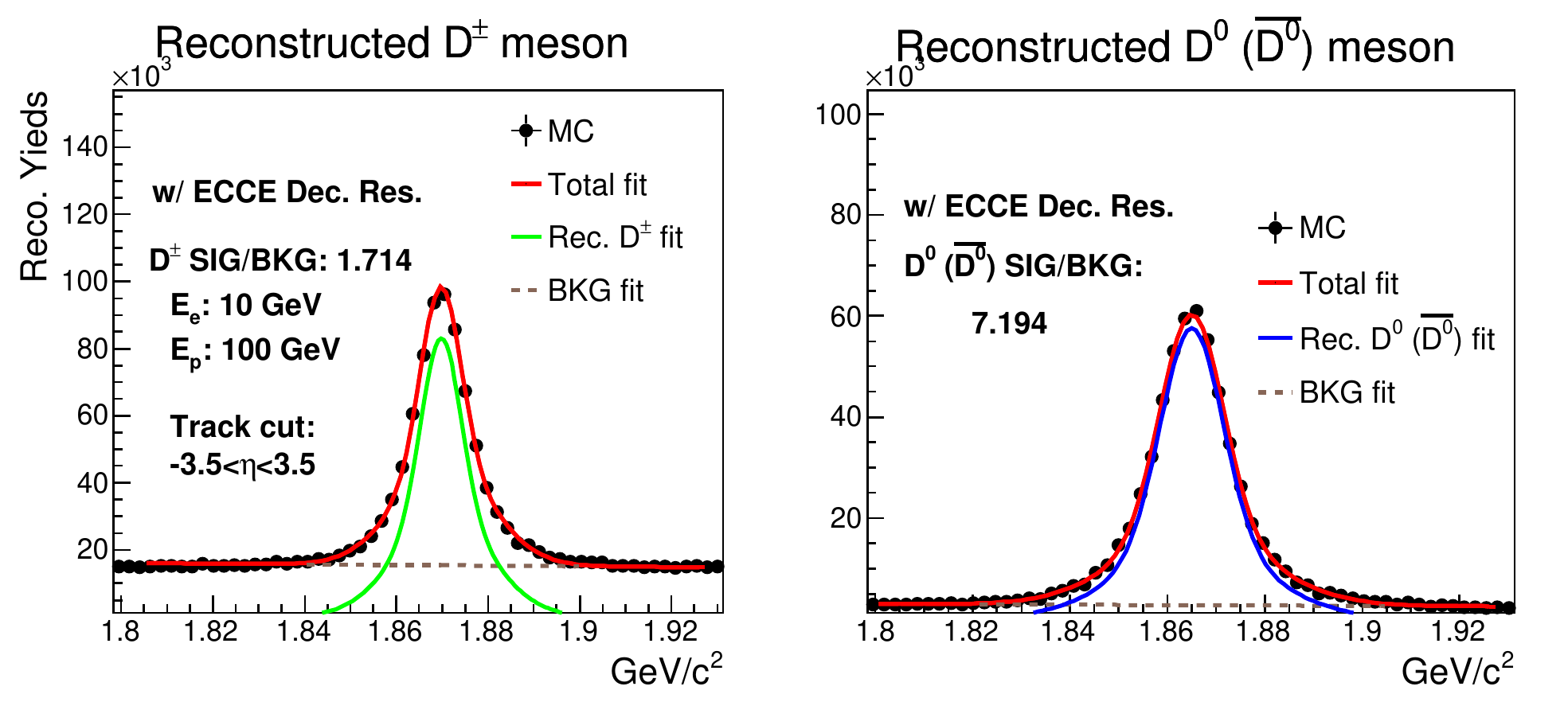}
\includegraphics[width=0.74\textwidth]{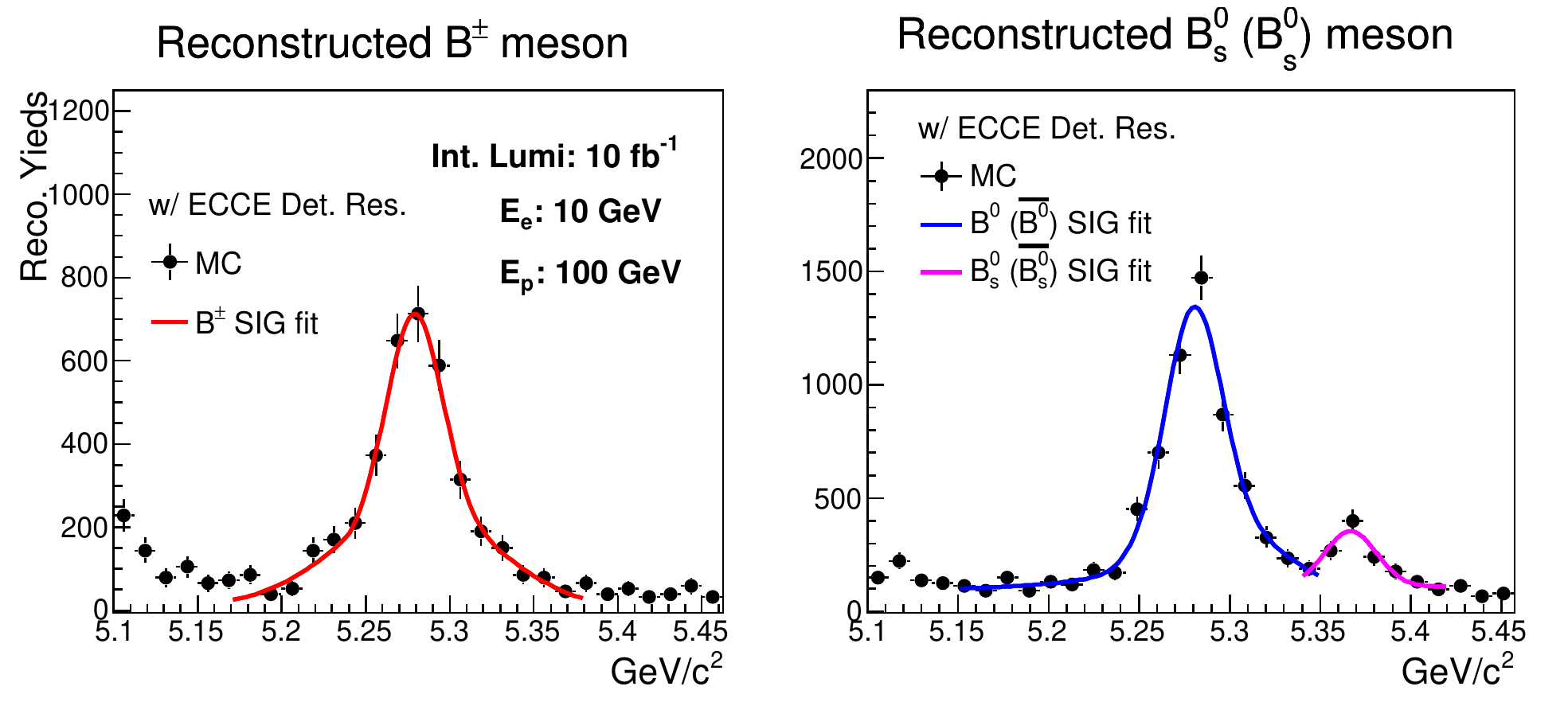}
\caption{Reconstructed $D^{\pm}$ ($D^{\pm} \rightarrow K^{\mp}\pi^{\pm}\pi^{\pm}$), $D^{0}$ ($\bar{D^{0}}$) ($D^{0} (\bar{D^{0}}) \rightarrow K^{\mp} \pi^{\pm}$), $B^{\pm}$ ($B^{\pm} \rightarrow J/\psi + K^{\pm}$), $B^{0}$ ($\bar{B^{0}}$) ($B^{0} (\bar{B^{0}}) \rightarrow  J/\psi + K^{\pm} + \pi^{\mp}$), and $B_{s}^{0}$ ($\bar{B_{s}^{0}}$) ($B_{s}^{0} (\bar{B_{s}^{0}}) \rightarrow J/\psi + \phi$) mass spectrums with the EIC project detector reference design in 10 GeV electron and 100 GeV proton collisions. The integrated luminosity used for the physics projection is 10 $fb^{-1}$.}
\label{ref_DB_reco}
\end{figure}

\subsection{Heavy flavor physics with the EIC project detector reference design}

A series of heavy flavor physics simulation studies \cite{lanl_hf1, lanl_hf2, ustc_hf} have been carried out with the detector performance of the EIC project detector reference design. The physics simulation framework consists of event generation in PYTHIA6/8 \cite{py6, py8}, smearing package of detector response extracted from the GEANT4 simulation, and particle finding and reconstruction algorithms. The electron and proton/nucleus beam has a crossing angle of 25 mrad at the EIC project detector Interaction Region (IP) to mitigate the related EIC background. This beam crossing angle introduced in the latest EIC accelerator design has been included in the simulation configuration. Fig.~\ref{ref_DB_reco} shows the mass distributions of reconstructed $D^{\pm}$, $D^{0}$ ($\bar{D^{0}}$), $B^{\pm}$, $B^{0}$ ($\bar{B^{0}}$), and $B_{s}^{0}$ ($\bar{B_{s}^{0}}$) with the detector performance of the EIC project detector reference design in 10 GeV electron and 100 GeV proton collisions \cite{lanl_hf2}. Due to the great precision of the displaced vertex and tracking momentum resolution of the EIC reference design (see Fig.~\ref{ref_trk_mom} and Fig.~\ref{ref_trk_dca}), Clear and pronounced signals can be obtained for these reconstructed D and B mesons with only one year of EIC operation. The extracted signal over background ratios are significantly better than those values measured in heavy ion collisions, which is due to the better detector performance and lower beam/collision backgrounds at EIC. In addition to open heavy flavor hadron reconstruction, heavy quarkonia such as $J/\psi$ \cite{ustc_hf} and heavy flavor jets \cite{lanl_hf1, lanl_hf2}, which have larger masses and a larger number of decay particles or constituents, have been successfully reconstructed or tagged with the EIC project detector reference design in electron+proton and electron+nucleus collisions. The unique kinematic coverage and high precision to be obtained by the future EIC heavy flavor hadron and jet measurements will provide significantly better sensitivities and new insights to explore the inner structure of nucleon and nucleus within the little constrained high and low Bjorken-x region and shed light on the mass/matter formation puzzle such as the hadronization mechanism. The precision of the proposed EIC heavy flavor measurements mainly relies on the high-granularity and low-material-budget EIC inner vertex and tracking detector based on the MAPS technology to provide precise primary and displaced vertex reconstruction and track momentum determination. The AC-LGAD based ToF detector plays an important role in identifying the PID information in the low momentum region for heavy flavor particle reconstruction at the EIC.

\begin{figure}[tbh!]
\centering
\includegraphics[width=0.85\textwidth]{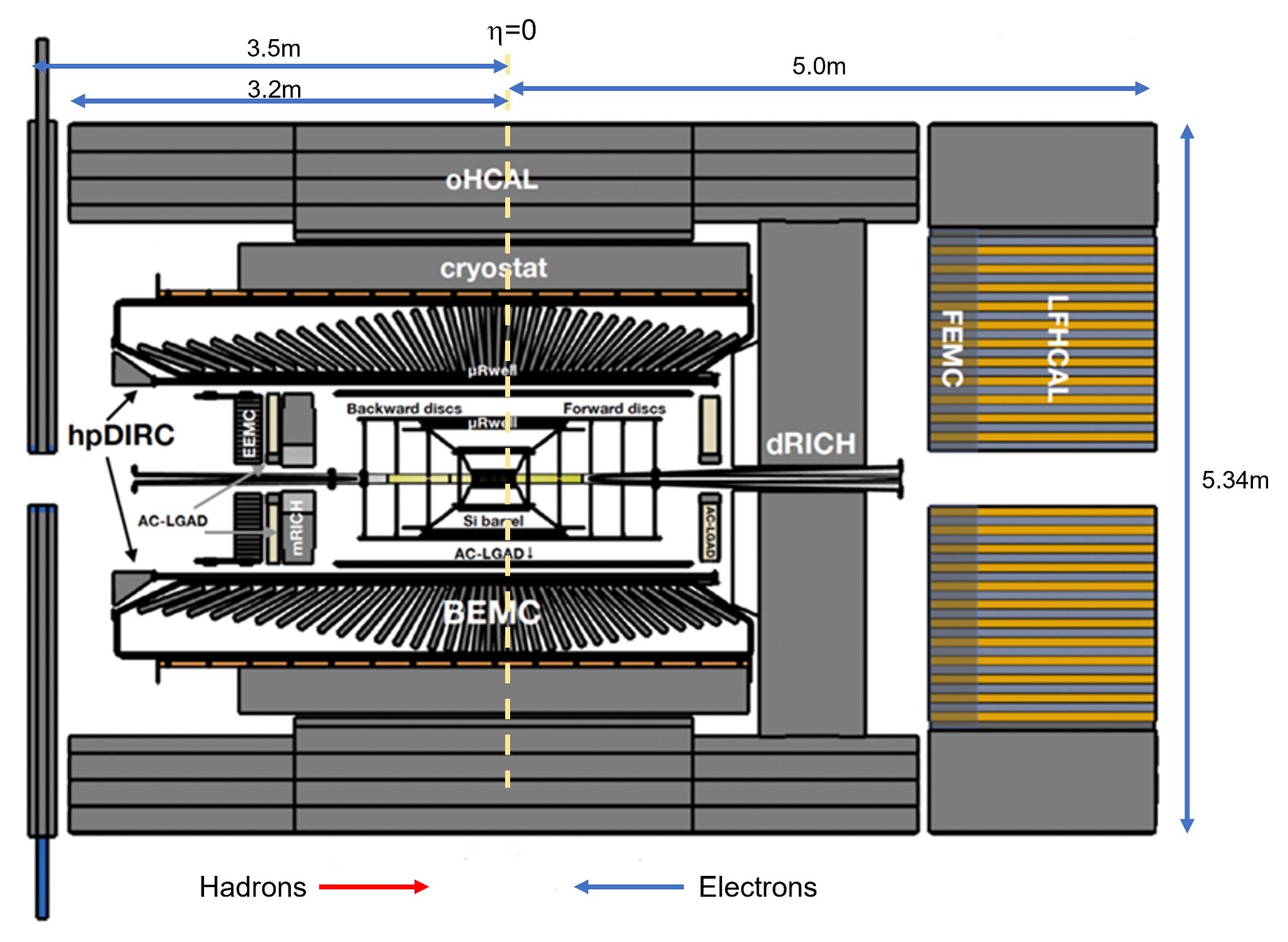}
\caption{Side view of the ePIC detector current design. The ePIC detector current design includes the tracking, PID, electromagnetic calorimeter and hadron calorimeter subsystems. The ePIC detector is 8.2 m long along the z axis and 5.34 m high along the y axis. It uses a new 1.7~T magnet with similar dimensions as the Babar magnet.}
\label{ePIC_dec}
\end{figure}

\begin{figure}[htb!]
\centering
\includegraphics[width=0.98\textwidth]{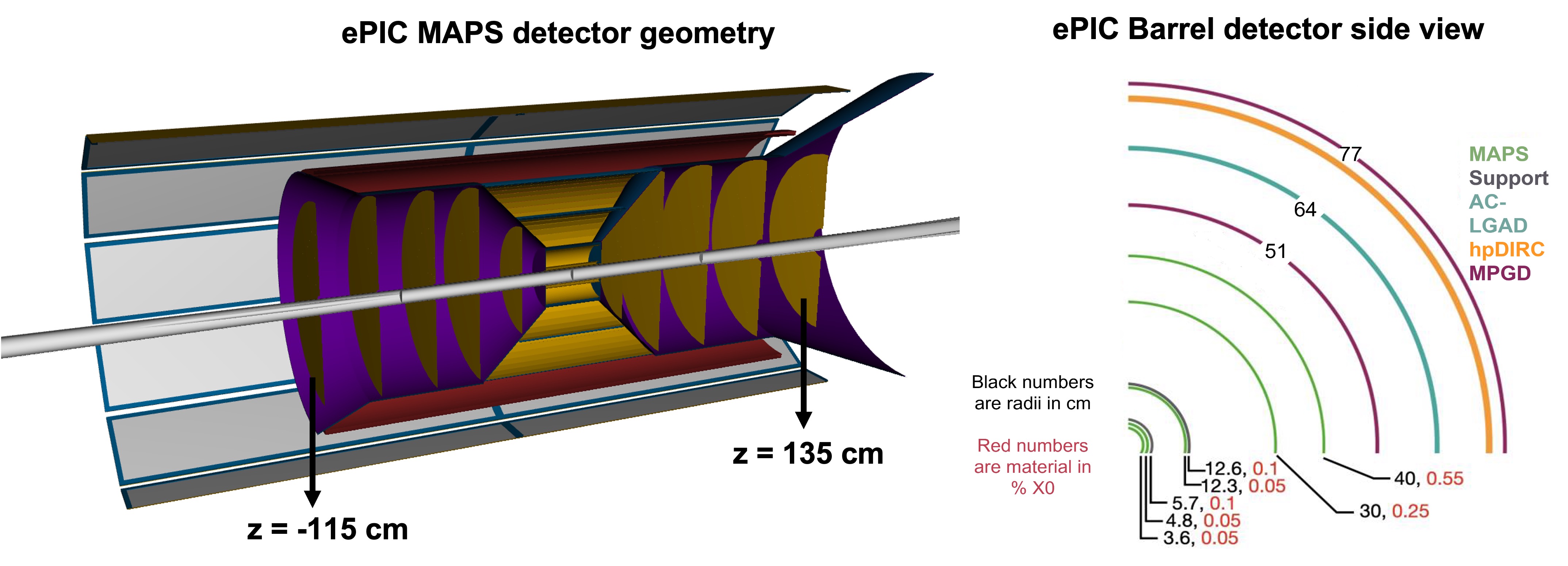}
\caption{Left: geometry of the current design of the ePIC MAPS vertex and tracking detector (golden) and its service cone (puple) implemented in simulation. Right: the side view of the ePIC barrel detector, in which includes the ePIC barrel MAPS vertex and tracking geometry parameters.}
\label{ePIC_tracking}
\end{figure}

\section{EIC silicon vertex and tracking subsystem technical design development}

After the EIC project detector reference design selection, a new detector collaboration: ePIC has been formed in July 2022 to lead the EIC project detector technical design based on the EIC reference design. The EIC project detector is expected to start operate in 2032, and the existing Babar magnet may not be fully functional during the EIC data collection period. Therefore a new magnet with the same dimension of the Babar magnet but slightly higher magnetic field at 1.7~T is introduced to be adapted by the ePIC detector design. Several detector geometry optimizations have been performed for the vertex and tracking subsystem, the PID subsystem and calorimeter subsystem of the current ePIC detector design. Fig.~\ref{ePIC_dec} shows the sideview of the updated ePIC detector geometry with a new magnet, which has similar dimension as the Babar magnet and its maximum magnetic field is 1.7~T. The ePIC detector is asymmetric along the z axis. The length in the hadron beam going direction from the primary vertex is 5.0~m long and the length in the electron beam going direction is 3.2~m. The height of the current ePIC detector design is 5.34 m. Following the EIC project detector reference design, the ePIC vertex and tracking detector still includes the MAPS inner vertex and tracking subsystem, the MPGD tracking subsystem and the AC-LGAD outer tracker. Fig.~\ref{ePIC_tracking} shows the geometry of the ePIC MAPS vertex and tracking detector and its service cone, which are implemented in simulation.

\begin{table}[tbh]
\caption{ePIC MAPS barrel vertex and tracking detector geometry}
\label{ePIC_barrel_trk}
\centering
\begin{tabular}{|clc|c|c|c|c|}
\hline
Index & $r$ (cm) & $z_{\rm min}$ (cm) & $z_{\rm max}$ (cm) & pixel pitch ($\mu m$) & Material Budget ($X/X_{0}$) \\  \hline
 1  & 3.6 & -13.5 & 13.5  &  10  & 0.05$\%$  \\ 
 2  & 4.8 & -13.5 & 13.5  &  10  & 0.05$\%$ \\
 3  & 12 & -13.5 & 13.5  &  10  & 0.05$\%$ \\ 
 4  & 27 & -27 & 27 & 10  & 0.25$\%$ \\ 
 5  & 42 & -42 & 42 & 10  & 0.55$\%$ \\
\hline
\end{tabular}
\end{table}

\begin{table}[tbh]
\caption{ePIC MAPS hadron endcap tracking detector geometry}
\label{ePIC_had_trk}
\centering
\begin{tabular}{|clc|c|c|c|c|}
\hline
Index & $z$ (cm) & $r_{\rm in}$ (cm) & $r_{\rm out}$ (cm) & pixel pitch ($\mu m$) & Material Budget ($X/X_{0}$) \\  \hline
 1  & 25 & 3.676 & 23  &  10  & 0.24$\%$  \\ 
 2  & 45 & 3.676 & 43  &  10  & 0.24$\%$ \\
 3  & 70 & 3.842 & 43  &  10  & 0.24$\%$ \\ 
 4  & 100 & 5.443 & 43  &  10  & 0.24$\%$ \\ 
 5  & 135 & 7.014 & 43  &  10  & 0.24$\%$ \\
\hline
\end{tabular}
\end{table}

\begin{table}[tbh]
\caption{ePIC MAPS electron endcap tracking detector geometry}
\label{ePIC_ele_trk}
\centering
\begin{tabular}{|clc|c|c|c|c|}
\hline
Index & $z$ (cm) & $r_{\rm in}$ (cm) & $r_{\rm out}$ (cm) & pixel pitch ($\mu m$) & Material Budget ($X/X_{0}$) \\  \hline
 1  & -25 & 3.676 & 23  &  10  & 0.24$\%$  \\ 
 2  & -45 & 3.676 & 43  &  10  & 0.24$\%$ \\
 3  & -65 & 3.676 & 43  &  10  & 0.24$\%$ \\ 
 4  & -90 & 4.006 & 43  &  10  & 0.24$\%$ \\ 
 5  & -115 & 4.635 & 43  &  10  & 0.24$\%$ \\
\hline
\end{tabular}
\end{table}

The optimized geometry parameters of the ePIC MAPS vertex and tracking detector are shown in Table~\ref{ePIC_barrel_trk}, Table~\ref{ePIC_had_trk} and Table~\ref{ePIC_ele_trk}. Following the EIC project detector reference design, the ePIC MAPS barrel vertex and tracking detector still has 5 layers, however the updated design covers a relatively larger active detector area and consists of a larger number of channels. The inner two vertex layers are moved to radii at 3.6 cm and 4.8 cm considering the EIC beam bake out requirement (5 mm clearance from the EIC beam pipe) and the new ITS3-like sensor size limitation \cite{its3, its3-d1, its3-d2}. The third vertex layer radius is changed to  12 cm to provide a good level arm for track reconstruction. As listed in Table~\ref{ePIC_barrel_trk}, the 4th and 5th MAPS barrel layers are moved further away to improve the track momentum and spatial resolutions. 

\begin{figure}[hbt]
\centering
\includegraphics[width=0.96\textwidth]{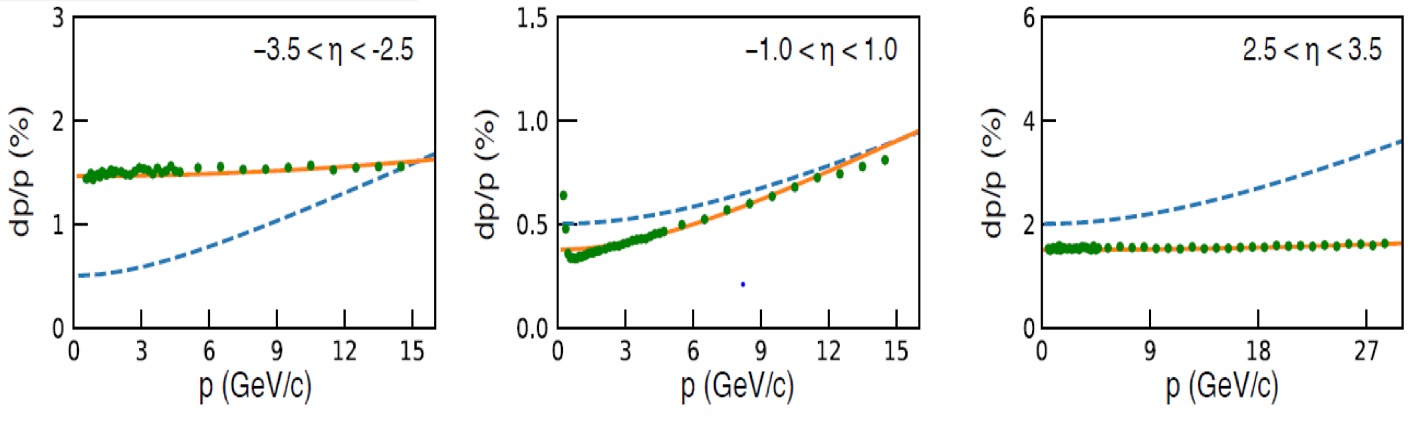}
\includegraphics[width=0.96\textwidth]{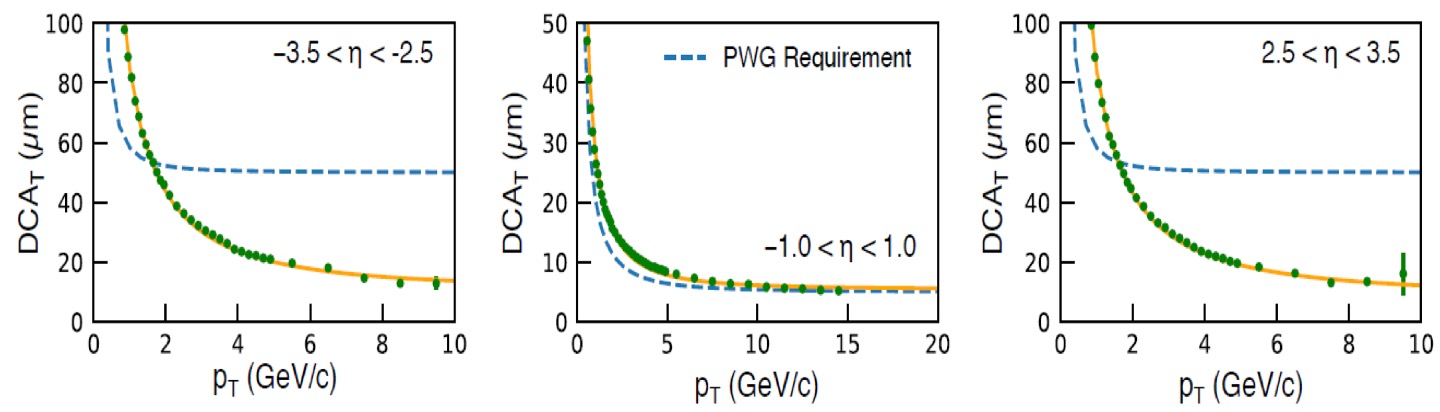}
\caption{Tracking performance of the current ePIC detector design in the pseudorapidity regions of $-3.5 < \eta < -2.5$, $-1.0 < \eta < 1.0$ and $2.5 < \eta < 3.5$. Track momentum dependent momentum resolution (top row) and transverse momentum dependent transverse Distance of Closest Approach ($DCA_{2D}$) resolution (bottom row) are evaluated with the new 1.7~T ePIC magnet. The EIC yellow report tracking requirements in the respective pseudorapidity regions are highlighted in blue dashed lines.}
\label{ePIC_trk_perform}
\end{figure}

The ePIC MAPS hadron endcap detector (geometry parameters listed in Table~\ref{ePIC_had_trk}) is composed of 5 disks with expanded z coverage from 25 cm to 135 cm. The inner radius of the hadron endcap disks varies from 3.67 cm to 7.01 cm and the outer radius varies from 23 cm to 43 cm. Compared to the EIC project detector reference design, the ePIC MAPS electron endcap detector (geometry parameters listed in Table~\ref{ePIC_ele_trk}) includes one more disk and has updated longitudinal z and radius r parameters to make full usage of the allocated space. The ePIC silicon vertex and tracking detector design aims to utilize the next generation 65 nm MAPS technology and the ePIC MAPS layers/disks assume the sensor pixel pitch is 10 $\mu m$. The ePIC collaboration plans to use the ITS-3 type bent sensor for the inner three barrel layers to obtain low material budgets and high spatial resolution. The MAPS outer two barrel layers and endcap disks will use new flat MAPS sensors with similar feature of the ITS-3 type sensor. Ongoing EIC detector R$\&$D focuses on developing the EIC MAPS sensor design, the silicon detector readout architecture, detector mechanical structure and service part material budget reduction.

In addition to the MAPS vertex and tracking detector, other subsystems of the ePIC detector design and the estimated service parts have been implemented in GEANT4 simulation. Fig.~\ref{ePIC_trk_perform} shows the tracking performance of the current ePIC detector design in GEANT4 simulation. With the optimized ePIC detector design, the track momentum dependent momentum resolution in the pseudorapidity region of $-1 < \eta < 1$ and $2.5 < \eta < 3.5$ meet the EIC yellow report detector requirement as shown in the top row of Fig.~\ref{ePIC_trk_perform}. Although the tracking momentum resolution of the current ePIC detector design in the pseudorapidity region of $-3.5 < \eta < -2.5$ is worse than the EIC yellow report detector requirement, further studies to reduce the detector material budgets in the electron endcap region and applying particle flow method with joint tracking and calorimeter information are underway. The transverse momentum dependent transverse Distance of Closest Approach ($DCA_{2D}$) resolution in most pseudorapidity regions meets the EIC yellow report requirements. As the EIC backgrounds such as the beam gas and synchrotron radiation background, are still under study in simulation, the ePIC tracking detector geometry may be updated to provide sufficient number of hits for tracking pattern recognition. The ePIC tracking performance will be further updated with new simulation samples implemented with EIC backgrounds.

\section{EIC silicon detector R$\&$D status}

The EIC project R$\&$D program, which focuses on the needs of detector components that are baselined in the project detector, ePIC, started in 2022 and is supported through the Brookhaven National Laboratory EIC center by the US Department of Energy.  For the ePIC silicon vertex and tracking detector, the 65 nm MAPS technology has been recommended. The AC-LGAD technology has been proposed to be applied for the ePIC barrel and hadron endcap outer tracker. Alternative silicon detector technologies have been considered such as the Depleted MAPS (DMAPS) technology \cite{malta} for the EIC detector. Existing advanced silicon detector R$\&$D \cite{malta_tes} can provide good reference for the EIC project detector R$\&$D. Progress of the 65 nm MAPS and AC-LGAD R$\&$D, which are supported by the EIC project detector R$\&$D programs, will be discussed in the following sections.
 
 \subsection{EIC MAPS detector R$\&$D}
 
The MAPS prototype sensor for the ePIC silicon vertex and tracking detector is still under design. The related technical knowledge and aspects have been learnt from the parallel R$\&$D of the 65 nm MAPS by the ALICE ITS3 project \cite{its3}. Three R$\&$D projects (eRD104, eRD11, eRD113) have been supported by the EIC project detector R$\&$D program for the 65 nm MAPS related detector design and R$\&$D. The near-term goals of these three MAPS R$\&$D projects include (1) providing technical solutions to adapt the ITS3 based mechanical and electrical characteristics to the nominal EIC vertex layer radii based on the available reticle sizes, and optimization of the bending and interconnection techniques for the resulting configuration; (2) delivering the sensor stave configuration for the 4th and 5th middle layers and disk configurations for the endcap regions; (3) optimizing the number of stitched sensor units into a module composed of an aluminum flex PCB and optimized number of sensors; and (4) performing a survey about the requirements for the large-area sensor design in a multiple sensor to a single readout chain scheme.

\begin{figure}[hbt]
\centering
\includegraphics[width=0.9\textwidth]{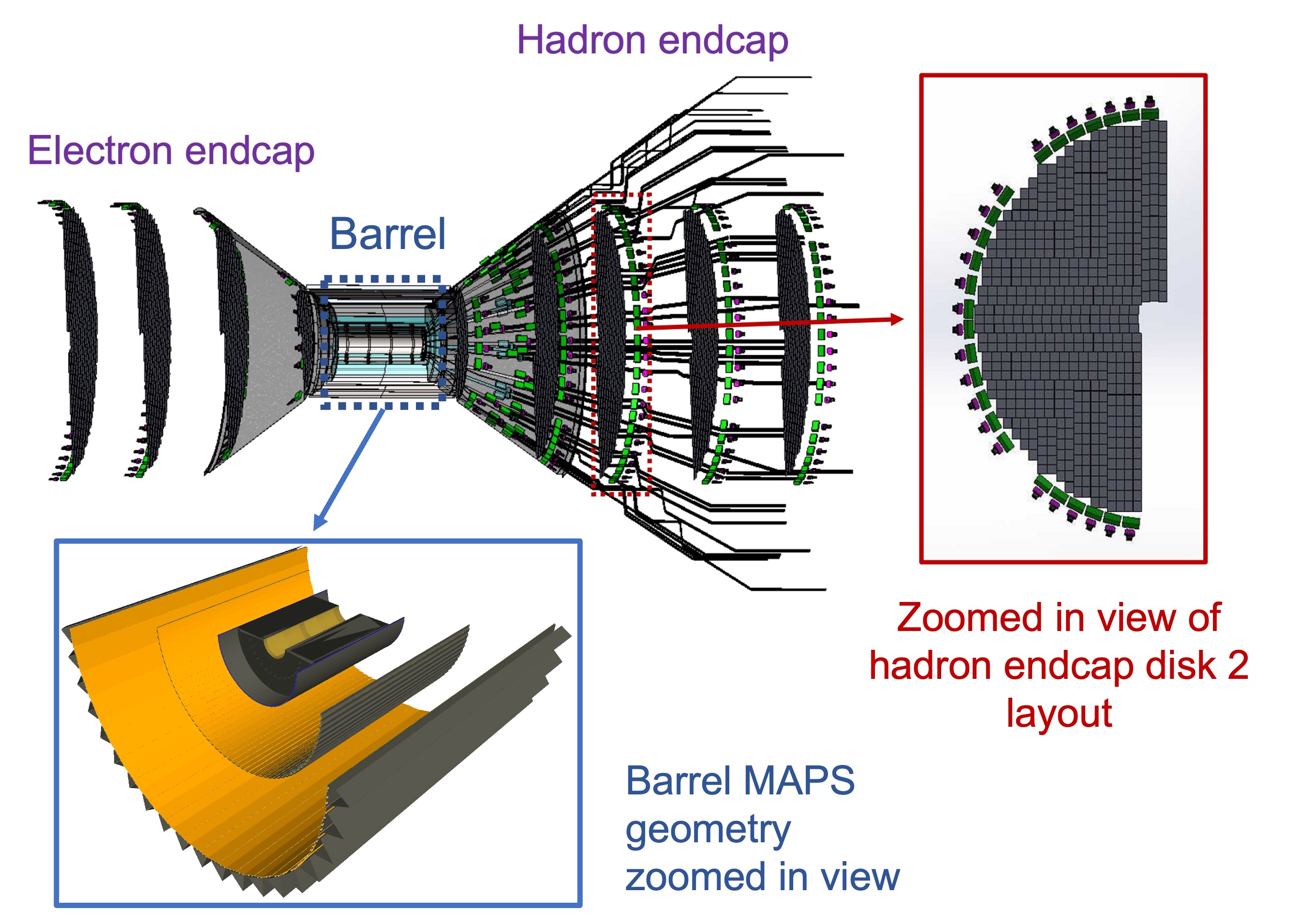}
\caption{The mechanical design of the MAPS vertex and tracking subsystem by the EIC eRD111 project according to the EIC project detector reference design. The hadron endcap disks locate on the right side and the electron endcap disks are in the left side. The current disk mechanical design utilizes the stitched MAPS sensors and the layout of the 3rd disk in the hadron endcap region is shown inside the red box.}
\label{eRD111_mec_design}
\end{figure}

The eRD104 project is investigating methods to significantly reduce the services load of the readout and power system for the EIC MAPS vertex and tracking detector. The eRD111 project focuses on developing the full detector mechanical design composed of the next generation 65 nm MAPS sensors. Fig.~\ref{eRD111_mec_design} shows the first version of the mechanical design of the MAPS vertex and tracking detector according to the EIC project detector reference design. The innermost three MAPS vertex layers are composed of curved MAPS sensors with a design similar to the proposed ITS3 sensor \cite{its3-d1, its3-d2}. The 4th and 5th MAPS barrel layers are made of MAPS sensor staves. As shown in Fig.~\ref{eRD111_mec_design}, the current ePIC MAPS detector mechanical design implements stitched MAPS sensors to construct its hadron and electron endcap disks. Besides the support structure of detector active volume, routines of the readout and power cables and the cooling system are included in the ePIC MAPS detector mechanical design. As the ePIC detector design is evolving, the silicon vertex and tracking detector mechanical design will be updated accordingly. A new EIC project detector R$\&$D project, eRD113, was recently set up to work on the EIC MAPS sensor design and associated sensor characterization. 

\subsection{EIC AC-LGAD detector R$\&$D}

\begin{figure}[hbt]
\centering
\includegraphics[width=0.92\textwidth]{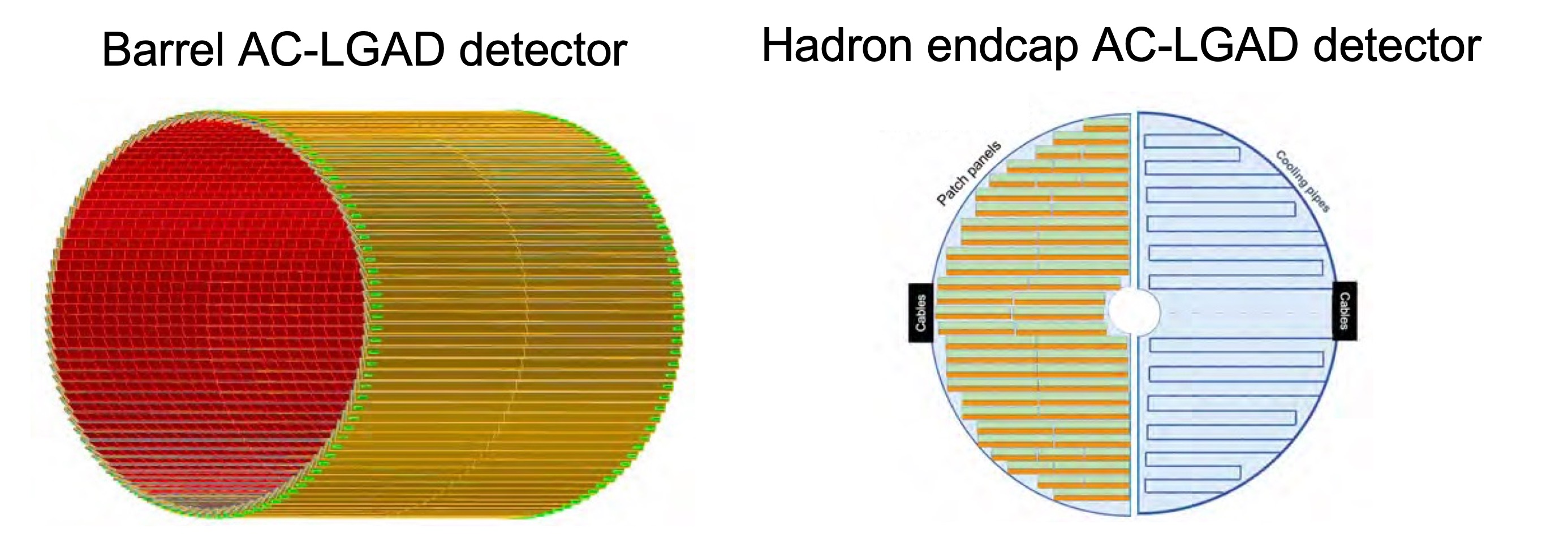}
\caption{The current design of the ePIC AC-LGAD barrel detector (left) and hadron endcap detector (right). In addition to the detector active volume layout, the associated readout module, support structure and cooling system have been implemented.}
\label{ePIC_AC_LGAD_design}
\end{figure}

The ePIC AC-LGAD barrel and hadron endcap detector design has been implemented with the support structure and cooling system as shown in Fig.~\ref{ePIC_AC_LGAD_design}. The ePIC barrel AC-LGAD tracker consists of around 2.4 million channels to cover around 10.9 $m^{2}$ area. The pixel size of the barrel AC-LGAD detector is 0.5~mm by 1.0~mm. The ePIC hadron endcap AC-LGAD tracker contains around 8.8 million pixels to cover around 2.22 $m^{2}$ area. The pixel size of the hadron endcap AC-LGAD detector is 0.5~mm by 0.5~mm. The AC-LGAD sensors are fabricated on thin high-resistivity silicon p-type substrates (around 50 $\mu$m thickness) and use a large and shallow n+ implant to cover the deep p+ layer for p-n junctions. Multiplication of the electrons traversing the device can reach from 5 to 100 and the current pulses at the terminals are created due to the drift holes through the substrate \cite{ac-lgad}. Moreover, the electrodes of the AC-LGAD sensors to connect to the readout electronics, are dilelectric metal pads separated from the n+ layer by a thin insulator. This feature allows adjacent pixels of the AC-LGAD sensor to share charges, which can improve the hit spatial resolution. The BNL group has fabricated both the 0.5~mm by 0.5~mm AC-LGAD pixel sensors and AC-LGAD strip sensors with different strip widths and pitches.
With the 0.5~mm by 1.0~mm strip sensor design, the ePIC barrel AC-LGAD tracker can achieve around 30~$\mu m$ spatial resolution in the $r\varphi$ plane and around 30~ps timing resolution. The ePIC hadron endcap AC-LGAD tracker can obtain around 30~$\mu m$ spatial resolution in the $xy$ plane and around 25~ps timing resolution with the 0.5~mm by 0.5~mm pixel sensor design.

\begin{figure}[hbt]
\centering
\includegraphics[width=0.97\textwidth]{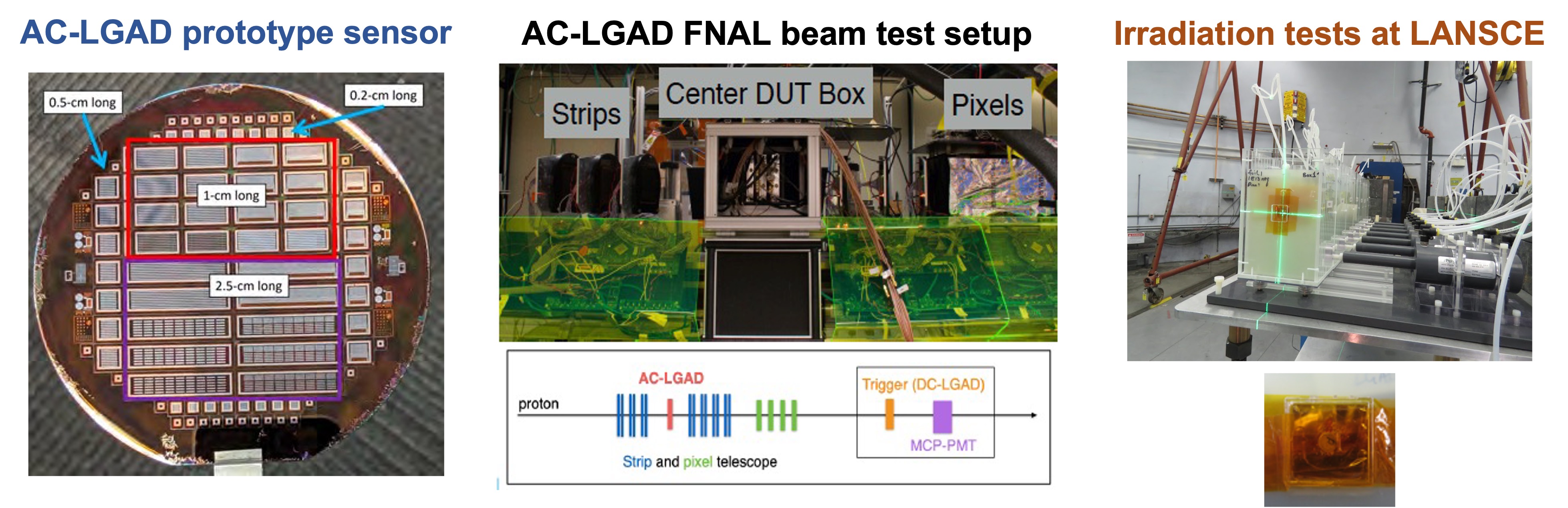}
\caption{The EIC AC-LGAD detector R$\&$D progress. New AC-LGAD pixel and strip prototype sensors have been produced at BNL and HPK (left). The AC-LGAD beam test setup at Fermi lab to evaluate its spatial and timing resolutions (middle). Irradiation test setup of the AC-LGAD prototype sensors at LANL LANSCE facility.}
\label{eRD112_AC_LGAD}
\end{figure}

A new EIC project detector R$\&$D project, eRD112, has been formed in 2022 to lead the ePIC AC-LGAD detector related R$\&$D. The project goals include (1) producing medium/large-area AC-LGAD sensors with different doping concentration, pitch and gap sizes between electrodes; and (2) providing a prototype ASIC design to readout AC-LGAD prototype sensors with around 30~ps timing resolution and low power consumption. Fig.~\ref{eRD112_AC_LGAD} shows the progress of the EIC AC-LGAD detector R$\&$D. New AC-LGAD prototype sensors with different pixel sizes have been produced at BNL and HPK. Beam tests of these new AC-LGAD sensors have been set up at Fermilab Test Beam Facility. From the beam tests, the AC-LGAD prototype sensors can achieve around 30~$\mu m$ spatial resolution and better than 30~ps timing resolution \cite{ac-lgad-2}. Although the EIC backgrounds are under evaluation, the radiation tolerance of new produced EIC silicon technologies should be characterized. Irradiation tests of EIC AC-LGAD prototype sensors have been performed at LANL LANSCE facility with 500~MeV proton beam. The irradiation doses delivered to these samples are $10^{13} - 10^{16}~n_{eq} cm ^{-2}$ and the radiation impacts on the performance of AC-LGAD sensors will be studied. 

\section{Summary and Outlook}

As the EIC project enters the new phase towards its realization, good progresses of the EIC project detector design and related R$\&$D have been achieved. The EIC project detector reference design includes the MAPS and AC-LGAD silicon vertex and tracking subsystems and its tracking performance can meet the EIC yellow report detector requirements in most kinematic regions. The newly formed ePIC collaboration is leading the EIC project detector technical design. The optimized ePIC detector design can achieve better tracking performance than the EIC project detector reference design. New EIC project detector R$\&$D
projects have been established to perform critical detector R$\&$D for the EIC project detector. The EIC MAPS and AC-LGAD prototype sensor design, detector R$\&$D and related detector mechanical design have obtained good progresses. New collaborators are encouraged to join the ePIC collaboration to deliver a low material budget and high precision silicon vertex and tracking detector for the EIC project.


\end{document}